\documentclass[twocolumn,showpacs,amsmath,amssymb]{revtex4}
\usepackage{graphicx}
\usepackage{dcolumn}
\usepackage{bm}
\usepackage[dvips]{epsfig}

\begin{document}
\title{{\bf Is Our Universe Decaying at an Astronomical Rate?}}
\author{Don N. Page}
\email{don@phys.ualberta.ca}
\affiliation{Theoretical Physics Institute, University of Alberta,
Edmonton, Alberta, Canada T6G 2G7}
\date{2006 Dec. 14}

\begin{abstract}

	Unless our universe is decaying at an astronomical rate (i.e.,
on the present cosmological timescale of Gigayears, rather than on the
quantum recurrence timescale of googolplexes), it would apparently
produce an infinite number of observers per comoving volume by thermal
or vacuum fluctuations (Boltzmann brains).  If the number of ordinary
observers per comoving volume is finite, this scenario seems to imply
zero likelihood for us to be ordinary observers and minuscule
likelihoods for our actual observations.  Hence, our observations
suggest that this scenario is incorrect and that perhaps our universe
is decaying at an astronomical rate.

\end{abstract}

\pacs{98.80.Qc, 04.60.-m, 95.30.Sf,  98.80.-k \hfill 
Alberta-Thy-19-06}

\maketitle

	``The most incomprehensible thing about the world is that it is
comprehensible,'' according to Einstein.  This mystery has both a
philosophical level \cite{theism} and a scientific level.  The
scientific level of the mystery is the question of how observers within
the universe have ordered observations and comprehensions of the
universe.

	If observers within the universe were sufficiently dominated
by those that are thermal or vacuum fluctuations (Boltzmann brains
\cite{Bolt,Rees,AS} or BBs), rather than by ordinary observers (OOs)
that arise from nonequilibrium processes like Darwinian evolution by
natural selection, then the probability would be very near unity that
a random observer would be such a BB.  However, the observations and
thoughts of such observers would be very unlikely to have the degree
of order we experience.  Therefore, if our observations can be
interpreted to be those of random observers, our ordered observations
would be statistical evidence against any theory in which BBs greatly
dominate.  A theory in which the universe lasts too long after a
finite period of OOs is in danger of being such a theory that is
statistically inconsistent with observation.

	This argument, a cosmic variant of the doomsday argument
\cite{C}, uses a version of the Copernican Principle or Vilenkin's
Principle of Mediocrity (PM) \cite{Vpm}, such as Bostrom's \cite{Bos}
{\it Strong Self-Sampling Assumption} (SSSA):  ``One should reason as
if one's present observer-moment were a random sample from the set of
all observer-moments in its reference class.''  This is similar to how
I might today state my {\it Conditional Aesthemic Principle} (CAP)
\cite{SQM}:  ``Unless one has compelling contrary evidence, one should
reason as if one's conscious perception were a random sample from the
set of all conscious perceptions.''  I would argue that the reference
class of all observer-moments (which I would call conscious
perceptions, each being all that one is consciously aware of at once)
should be the universal class of all observer-moments, but for the
present paper it is sufficient to take any reasonable restriction of
the reference class.

	For example, consider the reference class of scientists
observing the galaxy-galaxy correlation function (GGCF)
\cite{Hartle}.  This would almost certainly be only a very tiny
fraction of all observer-moments for BBs and presumably a much larger
(but still small) fraction of all observer-moments for OOs.  However,
if the BBs sufficiently dominate, most scientists observing the GGCF
would be BBs rather than OOs.  Then within this reference class, we
could compare the likelihoods that various theories give the observed
GGCF.  For BB theories that predict that almost all observer-moments
within this reference class are BBs, we would expect that almost all
of the ``observed'' GGCFs would be highly disordered, so the
likelihood of our observed GGCF would be very much smaller than in OO
theories in which most observer-moments within the same reference
class are OOs.  Therefore, unless we take the prior probability for BB
theories to be extremely near unity, our observed GGCF would be strong
statistical evidence that the posterior probability for BB theories
should be taken to be very small.

	In simple terms, our observations of order (e.g., a
comprehensible world) statistically rule out BB theories in which
observers formed by thermal or vacuum fluctuations greatly dominate
over ordinary observers.

	If we have a theory for a finite-sized universe that has
ordinary observers for only a finite period of time (e.g., during the
lifetime of stars and nearby planets where the ordinary observers
evolve), each of which makes only a finite number of observations
(perhaps mostly ordered), then the universe would have only a finite
number of ordinary observers with their largely ordered observations. 
On the other hand, if such a theory predicts that the universe lasts
for an infinite amount of time, then one would expect from vacuum
fluctuations an infinite number of observers (almost all extremely
short-lived, with very little ordered memory) and observations (almost
entirely with minuscule order or comprehension).  Such a BB theory
would give a much smaller likelihood of our actual observations than
an OO theory (in which BB's do not dominate over OOs) and so would be
discredited.

	Therefore, a good theory for a finite-sized universe apparently
should also predict that it have a finite lifetime.  (For example, this
was a property of the $k=+1$ Friedmann-Robertson-Walker dust model
universes.)

	For an infinite universe (infinite spatial volume in one
connected region, or an infinite number of connected regions in a
suitable multiverse picture of the universe \cite{Carr}), which is
predicted to be produced by eternal inflation \cite{Let,Vet,Set}, the
argument is not so clear.  In this case one could get an infinite
number of both ordinary observations and disordered observations, and
then there may be different ways of taking the ratio to give the
likelihoods of the possible observations within each reference class
\cite{LM,Vil,GSVW,ELM,VV,Bousso,BF,LindeBB,AGJ}.

	For example, one may \cite{Vil} define the probability $P_j$
for an observation to be the product $P_j = p_j f_j$ of the probability
$p_j$ for a particular pocket universe and of the probability $f_j$ of
the observation within that pocket.  Here I shall assume that $f_j$ is
regulated by taking the ratio of different kinds of observations within
a fixed comoving volume of the pocket universe.  This would give the
right answer for any finite bubble universe, no matter how large, but
of course it is an untested ansatz to apply this procedure to an open
or infinite bubble universe.

	With this assumption, the prospect of BB production then leads
one to conclude that any model universe should not last forever if it
has only a finite time period where ordinary observers dominate
\cite{lifetime}, though it must be noted that other regularization
schemes do not lead to this conclusion \cite{BF,Vil,LindeBB}.  However,
here we shall assume observations within a given pocket universe are
regulated by taking a finite comoving volume.

	The next question is what limits on the lifetime can be deduced
from this argument.  In \cite{lifetime} it was implicitly assumed that
the universe lasted for some definite time $t$ and then ended.  Then
the requirement was that the number of vacuum fluctuation observations
per comoving volume during that time not greatly exceed the number of
ordered observations during the finite time that ordinary observers
exist.  For any power-law expansion with exponent of order unity, and
using a conservative action of $10^{50}$ for a 1 kg brain to last 0.1
seconds, I predicted \cite{lifetime} that the universe would not last
past $t \sim e^{10^{50}}$ years, and for an universe that continues to
grow exponentially with a doubling time of the order of 10 Gyr, I
predicted that the universe would not last past about $10^{60}$ years. 
If instead one used what I now believe is a more realistic action of
$10^{42}$ for a 1 kg brain to be separated by 30 cm from the
corresponding antimatter \cite{HH-challenge,decay,42}, then the
corresponding times would be $t \sim e^{10^{42}}$ years and $10^{52}$
years respectively.

	Here I wish to emphasize that the expected lifetime should be
much shorter if the universe is expanding exponentially and just has a
certain decay rate for tunneling into oblivion.  Then the decay rate
should be sufficient to prevent the expectation value of the surviving
4-volume, per comoving 3-volume, from diverging and leading to an
infinite expectation value of vacuum fluctuation observations per
comoving 3-volume.  Since this minimum decay rate is set by the
asymptotic cosmological expansion rate of the universe, it may be said
to be astronomical (huge on the scale of the quantum recurrence time
\cite{GKS}.)

	Suppose \cite{decay} that the decay of the universe proceeds at
the rate, per 4-volume, of $A$ for the nucleation of a small bubble
that then expands at practically the speed of light, destroying
everything within the causal future of the bubble nucleation event. 
Then if one takes some event $p$ within the background spacetime, the
probability that the spacetime would have survived to that event is
$P(p) = e^{-A V_4(p)}$, where $V_4(p)$ is the spacetime 4-volume to the
past of the event $p$ in the background spacetime.  Then the
requirement that there not be an infinite expectation value of vacuum
fluctuation observations within a finite comoving 3-volume is the
requirement that the 4-volume within the comoving region, weighted by
the survival probability $P(p)$ for each point, be finite rather than
infinite.  For an asymptotically de Sitter background spacetime with
cosmological constant $\Lambda$, the expectation value of the 4-volume
of the surviving spacetime is finite if and only if \cite{decay}
\begin{equation} A > A_{\mathrm{min}} = \frac{9}{4\pi} H_\Lambda^4
 = \frac{\Lambda^2}{4\pi} \stackrel{>}{\sim} (20\ {\mathrm{Gyr}})^{-4}
 = e^{-562.5},
\label{eq:4}
\end{equation}
using $H_0 = 72\pm 8$ km/s/Mpc from the Hubble Space Telescope key
project \cite{Freedman} and $\Omega_\Lambda = 0.72\pm 0.04$ from the
third-year WMAP results of \cite{WMAP} to get $H_\Lambda =
H_0\sqrt{\Omega_\Lambda}$.

	For a universe that is spatially flat and has its energy
density dominated by the cosmological constant and by nonrelativistic
matter, as ours now seems to be, its $k=0$ FRW metric may be written
as
\begin{equation}
ds^2 = T^2[-d\tau^2 + (\sinh^{4/3}{\tau})(dr^2+r^2d\Omega^2)],
\label{eq:8}
\end{equation}
where $T = 2/(3 H_\Lambda)$.  This gives a survival probability
\begin{equation}
P(\tau) = \exp{\left[-\frac{16}{27}\frac{A}{A_{\mathrm{min}}}
          \int_0^\tau dx \sinh^2{x}
          \left(\int_x^\tau \frac{dy}{\sinh^{2/3}{y}}\right)^3\right]}
\label{eq:11}
\end{equation}
With the present earth population of nearly 7 billion, this would give
a minimal expected death rate of about 7 persons per century.  (Of
course, it could not be 7 persons in one century, but all 7 billion
with a probability of about one in a billion per century.)  It also
gives an upper limit on the present half-life of our universe of
\cite{decay} 19 Gyr.

	Although there are no observations that directly rule out the
suggestion here that our universe is decaying at an astronomical rate
(more likely than not to decay within 20 billion years), this
suggestion is surprising and would itself result in various
mysteries.  For example, it would leave it unexplained why the decay
rate $A$ is greater than $A_{\mathrm{min}} \approx (20\
{\mathrm{Gyr}})^{-4}$ and yet not so great to make it highly
improbable that our universe has lasted as long as it has.  There
might be a factor of 1000 window for the allowed $A$, but if, for
example, the decay rate is given by Eq. (5.20) of \cite{CDGKL}, this
translates into only a 0.6\% variation in the allowed value of the
gravitino mass \cite{decay}, which seems unnaturally restrictive. 
Furthermore, even if the annihilation rate $A$ can be within this
range for our part of the multiverse, it would still leave it
unexplained why it is not less than $A_{\mathrm{min}}$ in some other
part of the multiverse that also allows observers to be produced by
vacuum fluctuations.  If it were less in any such part of the
multiverse, then it would seem that that part would have an infinite
number of vacuum fluctuation observations that is in danger of
swamping the ordered observations in our part.

	Because of these potential problems with the prediction
suggested here (that the universe seems likely to decay within 20
billion years), one might ask how this prediction could be
circumvented.

	One obvious idea is that the current acceleration of the
universe is not due to a cosmological constant that would last forever
if the universe itself did not decay away.  Perhaps the current
acceleration is instead caused by the energy density of a scalar field
that is slowly rolling down a gentle slope of its potential
\cite{Linde,lifetime}.  However, this seems to raise its own issue of
fine tuning, since although the observership selection effect can
perhaps explain the small value of the potential, it does not seem to
give any obvious explanation of why the slope should also be small,
unless the scalar field is actually sitting at the bottom of a
potential minimum (effectively a cosmological constant).

	Another possibility is that the normalization employed in this
paper to get a finite number of ordinary observers, namely to restrict
to a finite comoving volume, might not be the correct procedure if our
universe really has infinite spatial volume.  This is indeed the
conclusion of several authors \cite{BF,Vil,LindeBB}.

	For example, Bousso and Freivogel \cite{BF} have recently
argued that the paradox does not arise in the local description
\cite{local} (unless the decay time is much, much longer) and is
evidence against the global description of the multiverse.  On the
other hand, both Linde \cite{LindeBB} and Vilenkin \cite{Vil} have more
recently given examples of how one may avoid the dreaded ``invasion of
Boltzmann brains'' within their global description. 

	Linde's solution \cite{LindeBB} is analogous to the following
situation \cite{decay}.  Consider imaginary humans who have a
`youthful' phase of 100 years of life with frequent and mostly ordered
observations, followed by a `senile' phase of trillions of years of
infrequent and mostly disordered observations.  Assume that the
trillions of years are sufficient to give many more total `senile'
disordered observations than `youthful' ordered observations for each
human.  One might think that most observations would then be
disordered, so that someone's having an ordered observation (which
would thus be very unlikely under this scenario) could count as
evidence against the theory giving this scenario.  However, if the
population growth rate of such humans is sufficiently high that at
each time the number of youthful humans and their ordered observations
outnumbers the senile humans and their disordered observations,
Linde's solution is that at each time the probability is higher that
an observation would be ordered than that it would be disordered.

	With the same aim, Vilenkin argues \cite{Vil} that since BBs
are equilibrium quantum fluctuation processes, they should be lumped
with bubbles in being calculated by the factor $p_j$ in $P_j = p_j f_j$
rather than by the factor $f_j$, which he suggests should be restricted
to observations formed by nonequilibrium processes.  Then he notes that
when BBs are counted in $p_j$, they are dominated by bio-friendly
bubbles that are also counted in $p_j$, since each such open bubble
gives an infinite universe and an infinite number of OOs.  He
concludes, ``As a result, freak observers [BBs] get a vanishing
relative weight.''

	I agree that these solutions are possible ways to regulate the
infinity of observations that occur in a universe that expands forever,
but they do not yet seem very natural \cite{AGJ}.  The problem arises
from the fact that if the youthful humans or OOs are always at late
times to outnumber the senile humans or BBs, the population of these
fictitious humans, or the volume of the universe in the original
example, must continue growing forever, producing an infinite number of
both youthful and senile fictitious humans or of both OOs and BBs in
cosmology.  Then it is ambiguous how one takes the ratio, which is the
fundamental problem with trying to solve the measure problem in
theories with eternal inflation.

	Of course, the fact that we have ordered observations and are
almost certainly not Boltzmann brains is strong evidence against what
I have here proposed as a natural way of using volume weighting in the
global viewpoint (unless the universe has a half-life less than 20
billion years \cite{decay}).  So in comparison with the observations,
my proposal of regulating by comoving volume is definitely worse than
the other prescriptions \cite{BF,Vil,LindeBB}, unless the universe
really is decaying at an astronomical rate.

	So if the prediction suggested in this paper (that our universe
seems likely to decay within 20 billion years) is wrong, it may be part
of our general lack of understanding of the measure in the multiverse. 
On the other hand, despite the fine-tuning problems mentioned above, it
is not obvious to me that it really is wrong, so one might want to take
it seriously unless and until some other really natural way is found to
avoid our ordered observations being swamped by disordered observations
from vacuum fluctuations.

	One might ask what the observable effects would be of the
decay of the universe, if ordinary observers like us could otherwise
survive for times long in comparison with 20 billion years.

	First of all, the destruction of the universe would likely
occur by a very thin bubble wall traveling extremely close to the speed
of light, so no one would be able to see it coming to dread the
imminent destruction.  Furthermore, the destruction of all we know (our
nearly flat spacetime, as well as all of its contents of particles and
fields) would happen so fast that there is not likely to be nearly
enough time for any signals of pain to reach our brains.  And no
grieving survivors will be left behind.  So in this way it would be the
most humanely possible execution.

	Furthermore, in an Everett many-worlds version of quantum
theory (e.g., \cite{SQM}), the universe will always persist in some
fraction of the Everett worlds (better, in some measure), but it is
just that the fraction or measure will decrease asymptotically toward
zero.  This means that there is always some positive measure for
observers to survive until any arbitrarily late fixed time, so one
could never absolutely rule out a decaying universe by observations at
any finite time, though sufficiently late ordinary observers would
have good statistical grounds for doubting the astronomical decay rate
suggested here.

	In any case, the decrease in the measure of the universe that
I am suggesting here takes such a long time that it should not cause
anyone to worry (except to try to find a solution to the huge
scientific mystery of the measure for the string landscape or other
multiverse theory).  However, it is interesting that the discovery of
the cosmic acceleration \cite{Ries,Perl} may not teach us that the
universe will certainly last much longer than the possible finite
lifetimes of $k=+1$ matter-dominated FRW models previously considered,
but it may instead have the implication that our universe is actually
decaying astronomically faster than what was previously considered.

	I have benefited greatly from numerous email discussions,
especially with Andrei Linde and Alex Vilenkin on eternal inflation
and with Jim Hartle and Mark Srednicki on probability, but also with
Andreas Albrecht, Anthony Aguirre, Nick Bostrom, Raphael Bousso,
Bernard Carr, Sean Carroll, David Coule, William Lane Craig, George
Ellis, Gary Gibbons, Steve Giddings, J. Richard Gott, Pavel
Krtou\v{s}, John Leslie, Don Marolf, Joe Polchinski, Martin Rees,
Michael Salem,  Glenn Starkman, Lenny Susskind, Neil Turok, and Bill
Unruh.  Financial support was provided by the Natural Sciences and
Engineering Research Council of Canada.


\begin{thebibliography}{99}
\bibitem{theism} A theistic explanation is that the universe was
created by an omniscient God who comprehends fully and who created
humans in His image to comprehend partially.
\end{thebibliography}
\end{document}